\documentclass[12pt]{article}

\usepackage[utf8]{inputenc}
\usepackage{hyperref}
\hypersetup{colorlinks}
\usepackage[dvips]{graphicx}
\usepackage{epsfig, amsmath, amssymb, verbatim, mathrsfs, array, layout, textcomp, latexsym, slashed, booktabs, color, mathtools, tikz, dsfont, cleveref, comment, enumerate, multirow, amsthm, xspace, physics, cite}

\numberwithin{equation}{section}

\theoremstyle{plain}
\newtheorem{thm}{Definition}

\newcommand{\FORM}{\textsc{Form}\xspace}
\newcommand{\CC}{\mathcal{C}}
\newcommand{\PP}{\mathcal{P}}
\newcommand{\CP}{\mathcal{CP}}
\newcommand{\scD}{\mathcal{D}}
\newcommand{\SU}[1]{\mathrm{SU}(#1)}
\newcommand{\SO}[1]{\mathrm{SO}(#1)}
\newcommand{\Uni}[1]{\mathrm{U}(#1)}
\newcommand{\diag}{\mathrm{diag}}

\begin{document}

\title{23, 381, 6242, 103268, 1743183, ...:\\ Hilbert series for CP-violating operators\\ in SMEFT\footnote{The title sequence is the total number of independent CP-violating operators, as defined in this paper, in the SMEFT with one fermion generation for mass dimensions 6, 8, 10, 12, 14, ... . }}

\author{Dan Kondo${}^{1}$\thanks{dan.kondo@ipmu.jp},
Hitoshi Murayama${}^{1,2,3}$\thanks{hitoshi@berkeley.edu, hitoshi.murayama@ipmu.jp, Hamamatsu Professor}, and
Risshin Okabe${}^{1}$\thanks{risshin.okabe@ipmu.jp}}
\date{\it\normalsize ${}^1$Kavli Institute for the Physics and Mathematics of the
  Universe (WPI), University of Tokyo, Kashiwa 277-8583, Japan\\
${}^2$Department of Physics, University of California\\ Berkeley, CA 94720, USA\\
${}^3$Theoretical Physics Group, Lawrence Berkeley National Laboratory\\
Berkeley, CA 94720, USA}

\maketitle

\begin{abstract}

We introduce a systematic method to classify the Standard Model Effective Field Theory (SMEFT) operators based on their CP properties with the Hilbert series techniques.
Our method makes it possible to enumerate operators violating CP symmetry in a few seconds.
We present the complete classification of dimension eight operators under CP transformation, 
and the number of CP-odd or CP-violating operators are listed up to dimension 14.
We also provide a companion code in \FORM that allows anybody to reproduce our results.

\end{abstract}

\newpage

\tableofcontents

\section{Introduction}

The Hilbert series has been used to systematically list operators in various effective field theories (EFTs) \cite{Jenkins:2009dy,Lehman:2015via,Henning:2015daa,lehman2016low,henning20172,Liao:2016qyd,Merlo:2016prs,Kobach:2017xkw,henning2017operator,Anisha:2019nzx,marinissen202083106786,graf20212,Yu:2021cco,Yu:2022nxj,Yu:2022ttm,Graf:2022rco}.
The introduction of the conformal group enabled the Hilbert series to remove the redundancies from equations of motion (EOM) and integration by parts (IBP) \cite{henning20172,henning2017operator}.
Furthermore, `folding' Dynkin diagrams was found to be useful to implement discrete symmetries into the Hilbert series; the Hilbert series with parity (P) and charge conjugation (C) were constructed in \cite{henning2017operator} and \cite{graf20212}, respectively.
However, both P and C have been applied only to theories in which each of them can be defined independently.
If P and C are not well-defined but CP is, we need to implement CP transformation in the Hilbert series.
The Standard Model is one example of such theories, and there has been no application of CP symmetry to the Hilbert series for the Standard Model Effective Field Theory (SMEFT).

The CP violation is important in any search for physics beyond the Standard Model. First of all, we know there must be more CP violations beyond the Cabibbo--Kobayashi--Maskawa phase to explain the baryon asymmetry of the universe. There are dedicated experimental searches underway in the $K$ system at \href{https://na62.web.cern.ch/Home/Home.html}{NA62} and \href{https://koto.kek.jp}{KOTO}, the $B$ system at \href{http://lhcb.web.cern.ch}{LHCb} and \href{https://www.belle2.org}{Belle II}, and neutrinos at \href{https://t2k-experiment.org}{T2K} and \href{https://novaexperiment.fnal.gov}{NO$\nu$A}. It is also the case that the sensitivity of experiments is higher for CP violation because the Standard-Model effects are usually highly suppressed. Therefore we can hope that even operators beyond mass dimension six can be probed experimentally.  This is why we believe it is important to list CP-violating operators in the SMEFT even for higher mass dimensions.

Previous research identified CP-odd operators in the SMEFT at dimension 6 \cite{alonso2014renormalization}, and listed bosonic CP-odd operators at dimension 8 \cite{remmen2019consistency}.
However, not all of these operators necessarily violate CP because they did not take ``rephasing" into account.
Rephasing is redefining the phases of the fields, which can remove some CP-violating phases.
A recent study \cite{bonnefoy2021beyond} found all CP-phases for dimension-six operators remaining after rephasing.

This paper shows how to construct CP-even and CP-odd Hilbert series for the SMEFT.
In \cref{sec:CC}, we introduce how to define charge conjugation for general compact Lie groups.
The charge conjugation for $\SU{N}$ has been discussed in \cite{arias2019discrete,henning2022outer}, and we improve their definition of charge conjugation.
In \cref{sec:CPHS}, we look into the implementation of CP transformation into the Hilbert series in detail.
We provide two definitions for operators that may violate CP symmetry depending on whether taking rephasing into account or not.
In \cref{sec:result}, we give the results of classifying dimension-eight operators and of counting operators up to dimension 14.
The results can be obtained using the accompanying \FORM code.
\Cref{sec:conclusion} is devoted to the conclusion. 
In addition, we clarify why charge conjugations that appear to involve complex conjugation are not anti-unitary but rather unitary operators in \Cref{sec:unitary_CC}. We also summarize the characters and Haar measures used in our explicit calculations in \Cref{sec:char} for the convenience of readers.

\section{Two types of charge conjugation}
\label{sec:CC}

We can think of charge conjugation as an automorphism that interchanges a representation with its complex conjugate representation.\footnote{The outer automorphisms for $\SO{4k}$ interchange one spinor with another spinor representation, but are not complex conjugation. The case of $\SO{8}$ allows for three such possibilities due to its triality. We will not discuss these cases since they are not relevant to the SMEFT.}
From this viewpoint, previous papers \cite{arias2019discrete,henning2022outer} have suggested that there are two inequivalent definitions of charge conjugation for the fundamental representation of $\SU{N}$ with even $N$.
However, we found that one of the two charge conjugations fails to be consistent with Hermitian conjugation on field operators.
In this section, it is shown that we can still define two types of charge conjugation on a general compact Lie group by weakening requirements in the previous studies.

\subsection{Requirements}
\label{sec:requiremnts}
Let us consider the charge conjugation $\CC$ which operates on some unitary representation of a compact Lie group $G$.
We suppose that charge conjugation $\CC$ interchanges fundamental and anti-fundamental representations.
We impose five requirements on $\CC$ to deduce its properties: 
\begin{enumerate}
  \item linearity,
  \item unitarity,
  \item $\CC^2=e^{i\theta}\mathds{1}$ with $\theta \in \mathbb{R}$,
  \item compatibility with $G$,
  \item consistency with Hermitian conjugation on field operators.
\end{enumerate}
We will see what each requirement means and discuss properties of $\CC$ derived from them in the following sections.

\subsection{Charge conjugation for fundamental representations}
\label{subsec:CC_complex}

First, we consider the charge conjugation on fundamental representations.
Let us denote the $N$-dimensional fundamental representation of a group $G$ by ${\bf N}$ 
and its complex conjugate representation by $\overline{\bf N}$.
The operation of charge conjugation $\CC$ is not closed within one representation itself because ${\bf N}$ and $\overline{\bf N}$ interchange among each other under charge conjugation.
Taking this into consideration, 
charge conjugation works only on the direct sum ${\bf N} \oplus \overline{\bf N}$.
For simplicity, we discuss a scalar field $\Phi=(\phi_1,\ \cdots,\ \phi_N)^T$ in the representation ${\bf N}$.
In this case, its Hermitian conjugate $\Phi^\dag=(\phi^\dag_1,\ \cdots,\ \phi^\dag_N)^T$ is in the representation $\overline{\bf N}$. Generalizations to fermion fields are straightforward.

The first requirement, linearity, restricts the form of $\CC$ as
\begin{align}
\CC \left( \begin{array}{c} \Phi \\ \Phi^\dag \end{array} \right)
= \left( \begin{array}{cc} 0 & C_{-}\\ C_{+} & 0 \end{array} \right)
\left( \begin{array}{c} \Phi \\ \Phi^\dag \end{array} \right) \,.
\label{eq:CC_NNbar}
\end{align}
Here, $C_{\pm}$ are $N \times N$ matrices.
Note that the symbol $\CC$ is used for both the group element of charge conjugation and its representation matrix.

Next, we will use the fifth requirement for the consistency of Hermitian conjugation on field operators. This requirement was missing in Refs.~\cite{arias2019discrete,henning2022outer}.
From \cref{eq:CC_NNbar}, the transformation laws under charge conjugation are
\begin{align}
    \Phi &\xrightarrow{\ \ \CC \ \ } C_- \Phi^\dag \, , 
    \label{eq:CC_Phi1}\\
    \Phi^\dag &\xrightarrow{\ \ \CC \ \ } C_+ \Phi \, .
    \label{eq:CC_Phidag1}
\end{align}
By taking Hermitian conjugation of \cref{eq:CC_Phi1}, we get another expression for transformation of $\Phi^\dag$:
\begin{equation}
     \Phi^\dag \xrightarrow{\ \ \CC \ \ } C^*_- \Phi \, . 
    \label{eq:CC_Phidag2}
\end{equation}
It should be noted that we considered Hermitian conjugation for quantum operators but not for matrices, which means the Hermitian conjugate of $C_-$ is $C_-^*$.
\Cref{eq:CC_Phidag1,eq:CC_Phidag2} implies 
\begin{equation}
    C_- = C^*_+.
\end{equation}
Therefore, we will use the notation $C_+=C$ and $C_- = C^*$.

The second requirement, unitarity, leads to
\begin{align}
\CC^{\dagger} \CC =
\left( \begin{array}{cc} 0 & C^\dag\\ C^T & 0 \end{array} \right)
\left( \begin{array}{cc} 0 & C^*\\ C & 0 \end{array} \right) 
=\left( \begin{array}{cc}  C^\dag C & 0 \\ 0 & C^T C^*  \end{array} \right)
= \mathds{1} \,,
\end{align}
and hence
\begin{align}
 C^\dag C = C C^\dag = \mathds{1} \,,
\end{align}
which means that the matrix $C$ is unitary. 

Let us use the third requirement; $\CC^2$ is a phase transformation of fields,\footnote{One might think that we need to interchange the first and second columns for the second $\CC$ operation because $\CC$ interchanges ${\bf N}$ and $\overline{\bf N}$. However, once the representation space is fixed, the representation matrix $\CC$ should not change its form according to what it operates on. Therefore, we must use the same representation matrices for two $\CC$s.}
\begin{align}
\CC^{2} =
\left( \begin{array}{cc} 0 & C^* \\ C & 0 \end{array} \right)
\left( \begin{array}{cc} 0 & C^* \\ C & 0 \end{array} \right) 
=\left( \begin{array}{cc}  C^* C & 0 \\ 0 & C C^*  \end{array} \right)
= e^{i\theta}\mathds{1}\, .
\end{align}
By comparing each block, we see that
\begin{align}
    C^* C = C C^* = e^{i\theta}\mathds{1} \,,
    \label{eq:CCstar}
\end{align}
and hence,
\begin{align}
    C = e^{i\theta} (C^*)^{-1} = e^{i\theta} C^T \,.
    \label{eq:CTandC1}
\end{align}
Taking transpose, we get
\begin{align}
    C^T= e^{i\theta}\, C \,.
    \label{eq:CTandC2}
\end{align}
From \cref{eq:CTandC1,eq:CTandC2}, we find
\begin{align}
    C = e^{i\theta}\, C^T = (e^{i\theta})^2\, C \,,
\end{align}
which means $e^{i\theta}=\pm 1$.
Furthermore, by taking determinant on the left and right-hand sides of \cref{eq:CTandC1}, we find
\begin{align}
\det C = (e^{i\theta})^{N} \det C^{T} = (e^{i\theta})^{N} \det C \,,
\end{align}
and hence $(e^{i\theta})^{N} = 1$.
For odd $N$, only $e^{i\theta}=1$ is allowed, while for even $N$, we have two possibilities $e^{i\theta}=\pm 1$.
Let us use $\CC_S$ for charge conjugation satisfying $\CC_S^2=+1$ and $\CC_A$ for $\CC_A^2=-1$.
The subscripts $S$ and $A$ indicate the symmetry of the matrix $C$: $C_S = C_S^T$ (symmetric) and $C_A = -C_A^T$ (anti-symmetric).

Finally, we will use the fourth requirement, the compatibility with $G$.
To see what this requirement means,  we define two discrete groups corresponding to $\CC_S$ and $\CC_A$ as
\begin{align}
    \Gamma_{\CC_S} &\equiv \{1,\ \CC_S\} \cong \mathbb{Z}_2\, , \\
    \Gamma_{\CC_A} &\equiv \{1,\ \CC_A,\ -1, -\CC_A\} \cong \mathbb{Z}_4\, .
\end{align}
The discrete group $\Gamma_\CC$ extends $G$ to the disconnected group $\widetilde{G}$:
\begin{equation}
    \widetilde{G} \equiv G\rtimes \Gamma_\CC \equiv \{g\gamma\ |\ g\in G,\ \gamma\in\Gamma_\CC\}\, .
\end{equation}
In order for $\widetilde{G}$ to be well-defined as a group, the product of two elements $g\gamma,g'\gamma'\in \widetilde{G}$ has to be in the form of $g''\gamma''\in \widetilde{G}$.
Thus, we have
\begin{equation}
    g\gamma \cdot g'\gamma' = (g\gamma g' \gamma^{-1})(\gamma\gamma') \equiv g''\gamma'' \,.
\end{equation}
Therefore, $\gamma g' \gamma^{-1}$ needs to be an element of $G$.\footnote{In other words, $\gamma$ is an automorphism of the group $\gamma: G \rightarrow G$ which preserves the multiplication rules (endomorphism). An inner automorphism is a conjugation by an element of $\gamma \in G$ and hence does not extend the group. What is interesting is an outer automorphism that does not belong to $G$.}
This gives our fourth requirement; $\CC g \CC^\dag$ is an element of $G$ for all $g\in G$.
Any element of $G$ can be represented on ${\bf N} \oplus \overline{\bf N}$ as
\begin{align}
U(g) \left( \begin{array}{c} \Phi \\ \Phi^\dag \end{array} \right)
= \left( \begin{array}{cc} e^{i T^{a} \omega^{a}} & 0 \\
	0 & V^{\dagger} e^{-i T^{aT} \omega^{a}} V \end{array} \right)
\left( \begin{array}{c} \Phi \\ \Phi^\dag \end{array} \right) \,.
\label{eq:U(g)}
\end{align}
Here $\{T^{a}\}$ are the unitary ${\bf N}$ representation of Lie algebra of $G$
and $\omega^{a} \in \mathbb{R}$ are real parameters. 
Its complex conjugate representation is equivalent to $\left(e^{i T^{a} \omega^{a}}\right)^* = e^{-i T^{aT} \omega^{a}}$ up to a unitary transformation $V$.
Without loss of generality, we can set $V=1$ by changing the basis for the $\overline{\bf N}$ representation. Then the element $\CC g \CC^\dag$ is represented as
\begin{align}
\CC U(g) \CC^\dag
&= \left( \begin{array}{cc} 0 & e^{-i\theta} C^{\dagger}\\ C & 0 \end{array} \right)
\left( \begin{array}{cc} e^{i T^{a} \omega^{a}} & 0 \\
	0 & e^{-i T^{aT} \omega^{a}} \end{array} \right)
\left( \begin{array}{cc} 0 & C^{\dagger}\\ e^{i\theta} C & 0 \end{array} \right)
\notag\\[3pt]
&= \left( \begin{array}{cc}
	C^{\dagger} e^{-i T^{aT} \omega^{a}} C & 0 \\
	0 & C e^{i T^{a} \omega^{a}} C^{\dagger}
	\end{array} \right) \,.
\label{eq:CgC}
\end{align}
At the same time, this is also an element of $G$, 
this can be written with other real parameters $\eta^a$ as
\begin{align}
\left( \begin{array}{cc}
	C^{\dagger} e^{-i T^{aT} \omega^{a}} C & 0 \\
	0 & C e^{i T^{a} \omega^{a}} C^{\dagger}
	\end{array} \right)
= \left( \begin{array}{cc}
	e^{i T^{a} \eta^{a}} & 0 \\
	0 & e^{-i T^{aT} \eta^{a}}
	\end{array} \right) \,.
\label{eq:compatibility}
\end{align}
Therefore, the lower right block of the right-hand side has to be a complex conjugate of its upper left block:
\begin{equation}
    C e^{i T^{a} \omega^{a}} C^{\dagger} = (C^{\dagger} e^{-i T^{aT} \omega^{a}} C)^*\, ,
\end{equation}
and hence
\begin{equation}
    C^{*} C e^{i T^{a} \omega^{a}} = e^{i T^{a} \omega^{a}} C^{*} C \,.
\end{equation}
That is, $C^{*} C$ commutes with any element of $G$.
By Schur's lemma, $C^{*} C$ is proportional to an identity matrix
\begin{align}
    C^{*} C = \alpha\,\mathds{1} \,.
\end{align}
This is exactly the same form as \cref{eq:CCstar}. Therefore we find
\begin{equation}
    \alpha = e^{i\theta} = \left\{
        \begin{aligned}
            & +1 && \text{for odd $N$} \\
            & \pm 1 && \text{for even $N$}
        \end{aligned}
    \right. \, .
\end{equation}

To summarize,
\begin{itemize}
  \item for odd $N$, charge conjugation has to be symmetric and involutive: 
    \[
        C_S^T=C_S \text{\ \ and \ } \CC_S^2=+\mathds{1}.
    \]
  \item for even $N$, there are two possibilities.
    \begin{itemize}
      \item symmetric: \ $C_S^T=C_S$ \ and \ $\CC_S^2=+\mathds{1}$,
      \item anti-symmetric: \ $C_A^T=-C_A$ \ and \ $\CC_A^2=-\mathds{1}$.
    \end{itemize}
\end{itemize}

Let us revisit the transformation laws of $\Phi$ and $\Phi^\dag$:
\begin{align}
  \Phi & \xrightarrow{\ \ \CC \ \ } C^* \Phi^\dag\, , 
  \label{eq:CC_Phi2}\\
  \Phi^\dag & \xrightarrow{\ \ \CC \ \ } C \Phi\, .
  \label{eq:CC_Phidag3}
\end{align}
For $G=\Uni{1}$ in particular, $C$ is not a matrix but just a phase.
This phase is sometimes called ``charge-conjugation parity'' and assigned the symbol $\xi$ in \cite{weinberg_1995}.
\Cref{eq:CC_Phi2,eq:CC_Phidag3} seem to suggest that the charge conjugation $\CC$ is an anti-unitary operator, but we will show that unitary charge conjugation can be defined by considering its action on creation and annihilation operators in \cref{sec:unitary_CC}.

Previous papers \cite{arias2019discrete,henning2022outer} require charge conjugation $\CC$ to be (1) linear, (2) unitary, (3) involutive ($\CC^2=+1$), and (4) consistent with $\SU{N}$.
In a similar way we did, they prove that there are two versions of charge conjugation for even $N$: symmetric ($C_S^T=C_S$) and anti-symmetric ($C_A^T=-C_A$).
However, when we additionally impose the fifth requirement in \cref{sec:requiremnts}, which is necessary in the context of QFT, $\CC_A$ fails to satisfy their third requirement, $\CC_A^2=+1$.
However, we can define $\CC_A$ consistently by weakening their third requirement; $\CC^2$ is not necessarily $+1$ but just a phase factor $e^{i\theta}$.

In this section, we have discussed charge conjugation only for fundamental representations.
If we want to know the symmetry properties of other general representations, we need to consider the action of charge conjugation onto the tensor products of fundamental and anti-fundamental representations. 
For example, charge conjugation of the adjoint representation can be understood by the transformation laws of fundamental and anti-fundamental representations.
Let us use the upper indices for fundamental representations and the lower indices for anti-fundamental representations as the tensor notation.
The transformation of the adjoint representation can be written as 
\begin{equation}
    {A^i}_j \xrightarrow{\ \ \CC \ \ } \eta_C (C^*)^{ik} C_{jl} {A^l}_k \, ,
\end{equation}
which is equivalent to
\begin{equation}
    A \xrightarrow{\ \ \CC \ \ } \eta_C C^* A^T C^T = \eta_C C^\dag A^T C \, .
    \label{eq:CC_adj}
\end{equation}
Since the adjoint representation is real, the overall phase factor $\eta_C$ is limited to $\pm 1$.
Furthermore, when $A$ is a gauge field as $A = A^a T^a$,
$\eta_C$ must be $-1$ to maintain the Lie algebra.

If we require $- C^\dag H^{aT} C = - H^a$ for the Cartan subalgebra $H^a$ as in \cite{grimus1997automorphisms}, the charge conjugation matrix is restricted to the symmetric one $C_S$.
This requirement comes from a viewpoint that charge conjugation should reverse all quantum numbers or, in other words, the diagonal components of the Cartan subalgebra.
However, we do not require this condition because such quantum numbers are not directly related to the ``charges" of particles in general cases. In addition, $\CC_A$ is much more natural for some cases from the point of view of the Hilbert series as we will see later.
Therefore, we will consider both $\CC_S$ and $\CC_A$ henceforth.

\subsection{Charge conjugation for (pseudo)real representations}
\label{subsec:real_rep}

For a (pseudo)real representation of Lie algebra $\{T^a\}$,
there is a unitary matrix $R$ such that 
\begin{equation}
    R^\dag (-{T^a}^T) R = T^a\, ,
\end{equation}
where the matrix $R$ is symmetric for real representations and anti-symmetric for pseudoreal representations.
In this case, the charge conjugation is an inner automorphism
\cite{georgi1999lie}.

One can define charge conjugation for (pseudo)real representations as a transformation from a representation to the same representation.
This definition is accomplished by setting the matrix $C$, which is defined in the previous section, to be equal to $R$.
In this case, fields in real representations are transformed by $\CC_S$ and those in pseudoreal representations are transformed by $\CC_A$.
For example, the fundamental representation ${\bf N}$ of $\mathrm{SO}(N)$ is real. Therefore, as long as the charge conjugation $\CC$ is defined for the irreducible representation ${\bf N}$, charge conjugation for ${\bf N}$ is limited to $\CC_S$.\footnote{Note that there are outer automorphisms that can act non-trivially on real representations, such as the ``parity'' element of $\mathrm{O}(2k)$. We do not discuss it here because it is not important for the SMEFT, but the interested readers are referred to \cite{henning2022outer}.}
In contrast, since the fundamental representation of $\mathrm{Sp}(2N)$ is pseudoreal, charge conjugation for this representation have to be $\CC_A$.

However, for a pseudoreal representation, if we allow $\CC$ to transform ${\bf N}$ to $\overline{{\bf N}}$ ({\it i.e.} if we let $\CC$ act on a reducible representation ${\bf N} \oplus \overline{{\bf N}}$), $\CC$ can be defined as $\CC_S$.
Note that, although this is possible, it would require an artificial differentiation of ${\bf N}$ and $\overline{\bf N}$ even when they are unitary equivalent.
Let us see an example of the Higgs doublet $H$.
This is a fundamental representation of  $\SU{2}$, which is a pseudoreal representation.
There are two definitions of charge conjugation $\CC_S$ and $\CC_A$ for $H$: 
\begin{equation}
  H = \left(\begin{array}{c} \phi^+ \\ \phi^0 \end{array}\right)
    \xrightarrow{\ \ \CC_S \ \ }
    \left(\begin{array}{c} \phi^{+\dag} \\ \phi^{0\dag} \end{array}\right)
    \xrightarrow{\ \ \CC_S \ \ }
    \left(\begin{array}{c} \phi^+ \\ \phi^0 \end{array}\right) = +H,
\end{equation}
\begin{equation}
  H = \left(\begin{array}{c} \phi^+ \\ \phi^0 \end{array}\right)
    \xrightarrow{\ \ \CC_A \ \ }
    \left(\begin{array}{c} \phi^{0\dag} \\ -\phi^{+\dag} \end{array}\right)
    \xrightarrow{\ \ \CC_A \ \ }
    \left(\begin{array}{c} -\phi^+ \\ -\phi^0 \end{array}\right) = -H.
    \label{eq:H_Ca}
\end{equation}
When we write the representation of $H$ as ${\bf 2}$,
the representation of $\CC_S H$ should be written as $\overline{\bf 2}$, and that of $\CC_A H$ is ${\bf 2}$ as is expected.

\subsection{Symmetric and anti-symmetric charge conjugations}
For even-dimensional representations, we can define charge conjugation in two ways, $\CC_S$ and $\CC_A$.
These symmetry properties are independent of the choice of basis.
When we change the basis of ${\bf N}$-representation field $\Phi$ by a unitary transformation $\Phi\to U\Phi$, the charge conjugation matrix $\CC$ changes as
\begin{align}
    \CC \to \CC' &=
    \left( \begin{array}{cc} U & 0\\ 0 & U^* \end{array} \right) 
    \left( \begin{array}{cc} 0 & C^*\\ C & 0 \end{array} \right)
    \left( \begin{array}{cc} U^\dagger & 0\\ 0 & U^T \end{array} \right) \notag\\
    &= \left( \begin{array}{cc} 0 & U C^* U^T \\ U^* C U^\dag & 0 \end{array} \right).
    \label{eq:basis}
\end{align}
The sign of the square of charge conjugation does not change by this transformation:
\begin{equation}
    {\CC'}^2 = \CC^2.
\end{equation}
Therefore, $\CC_S$ and $\CC_A$ cannot be mixed together by the change of the basis.

Similarly, $\CC_S$ and $\CC_A$ are gauge inequivalent. This is shown by replacing $U$ in \cref{eq:basis} with a group element $\exp(iT^a\omega^a)\in G$.
However, this gauge inequivalence does not mean that the extended groups by $\CC_S$ and $\CC_A$ are inequivalent.
Let us consider the case $G=\SU{2N}$.
If we take the choice 
\begin{equation}
    C_S = \mathds{1} = \mqty(\dmat{1, \ddots, 1}) \, , \quad 
    C_A = J \equiv \mqty(\dmat{\mqty{0 & 1\\ -1 & 0}, \ddots, \mqty{0 & 1\\ -1 & 0}}) \, ,
\end{equation}
there is a group element $-J \in \SU{2N}$, which satisfies
\begin{equation}
    \mqty(0 & C_S^*\\ C_S & 0) = \mqty(-J & 0 \\ 0 & -J^*) \mqty(0 & C_A^* \\ C_A & 0)\, .
\end{equation}
Therefore, the extended groups are equivalent:\footnote{This was shown very recently in \cite{Bischer:2022rvf}.}
\begin{equation}
    \SU{2N} \rtimes \Gamma_{\CC_S} = \SU{2N} \rtimes \Gamma_{\CC_A}\, .
    \label{eq:equivalence}
\end{equation}
In such a case, charge conjugation $\CC_S$ and $\CC_A$ result in the same physics.\footnote{For parity in $\SO{N}$, two extended groups are inequivalent, Pin$_+(N)$ and Pin$_-(N)$.}

\section{CP violation in the SMEFT}
\label{sec:CPHS}

\subsection{CP violation at dimension 4 or less}
\label{sec:CPV_SM}

The SM has two CP-violating sources: the phase $\delta_{\text{CKM}}$ in the Cabibbo--Kobayashi--Maskawa (CKM) matrix and the strong CP phase $\theta_{\text{QCD}}$.
We will briefly review the origins of the two phases and list all U(1) transformations that keep these phases unchanged.
Here we assume that neutrinos are massless up to mass dimension 4 (namely the absence of right-handed neutrinos) and the number of the fermion generations is three.

Let us begin with the $\SU{2}_W$ doublets and singlets in the quark sector
\begin{equation}
    Q_i^0=\begin{pmatrix} u_{iL}^0 \\ d_{iL}^0 \end{pmatrix},\ 
    u_{iR}^0,\ d_{iR}^0 \quad (i=1,2,3)\, .
\end{equation}
The upper script 0 means that they are weak eigenstates.
In terms of them, we can write the Yukawa interaction terms
\begin{equation}
    \mathcal{L}_Y = - (f_d)_{ij}\overline{Q_i^0} H d_{jR}^0 - (f_u)_{ij}\overline{Q_i^0} \tilde{H} u_{jR}^0 + \text{h.c.}\, .
    \label{eq:yukawa}
\end{equation}
Here, $H$ is the Higgs doublet and $\tilde{H} = i\sigma^2 H^*$ where $\sigma^I$ are the Pauli matrices.
The matrices $f_{u,d}$ are $3\times 3$ complex matrices.
By defining the vectors
\begin{equation}
    U_L^0 = \begin{pmatrix} u_{1L}^0 \\ u_{2L}^0 \\ u_{3L}^0 \end{pmatrix},\ 
    D_L^0 = \begin{pmatrix} d_{1L}^0 \\ d_{2L}^0 \\ d_{3L}^0 \end{pmatrix},\ 
    U_R^0 = \begin{pmatrix} u_{1R}^0 \\ u_{2R}^0 \\ u_{3R}^0 \end{pmatrix},\ 
    D_R^0 = \begin{pmatrix} d_{1R}^0 \\ d_{2R}^0 \\ d_{3R}^0 \end{pmatrix},
\end{equation}
we can rewrite \cref{eq:yukawa} as
\begin{equation}
    \mathcal{L}_Y = - (\overline{U_L^0} \ \overline{D_L^0}) H f_d D_R^0
        - (\overline{U_L^0} \ \overline{D_L^0}) \tilde{H} f_u U_R^0 + \text{h.c.}\, .
    \label{eq:yukawa2}
\end{equation}
When $H$ has the vacuum expectation value $\expval{H}=(0, v/\sqrt{2})^T$ with $v\in \mathbb{R}$, 
\cref{eq:yukawa2} becomes
\begin{equation}
    \mathcal{L}_Y = - \overline{D_L^0} M_d^0 D_R^0 - \overline{U_L^0} M_u^0 U_R^0 + \text{h.c.}\, ,
\end{equation}
where $M_{u,d}^0 = f_{u,d}v/\sqrt{2}$ are generally non-diagonal complex matrices.

Let us move to the mass eigenstates $U_{L,R}$ and $D_{L,R}$ by unitary transformations
\begin{equation}
    U_L^0 = V_{uL}U_L\, ,\ U_R^0 = V_{uR}U_R\, ,\ D_L^0 = V_{dL}D_L\, ,\ D_R^0 = V_{dR}D_R\, . 
\end{equation}
The unitary matrices $V$ diagonalize $M_{u,d}^0$ by bi-unitary transformations:
\begin{align}
    &V_{dL}^\dag M_d^0 V_{dR} = M_d = \diag(m_d,m_s,m_b)\, , \\
    &V_{uL}^\dag M_u^0 V_{uR} = M_u = \diag(m_u,m_c,m_t)\, .
\end{align}
If all quark masses are not degenerate, 
further unitary transformations like $U_L \to V'_{uL} U_L$ that keep the mass matrices invariant are limited to the vector-like and diagonal ones:
\begin{equation}
    V_{uL}^{\prime\dag} M_u V'_{uR} = M_u \quad \text{only if} \quad V'_{uL}=V'_{uR}=\diag(e^{i\theta_u}, e^{i\theta_c}, e^{i\theta_t})\, .
\end{equation}
The same goes for the down sector.
These U(1) transformations for each flavor are often called ``rephasing".

The degree of freedom of rephasing can be used to remove some of the complex phases in the CKM matrix $V_{\text{CKM}}$, which is defined in the charged current by
\begin{equation}
    J^+_\mu = \frac{g}{\sqrt{2}} \overline{U_L^0} \gamma^\mu D_L^0 = \frac{g}{\sqrt{2}} \overline{U_L} V_{\text{CKM}} \gamma^\mu D_L, \quad V_\text{CKM} = V^\dag_{uL}V_{dL}\, .
    \label{eq:CKM}
\end{equation}
As a result, we can remove all complex phases except one phase $\delta_\text{CKM}$ by using five rephasing out of six.
The only one remaining U(1) symmetry is $\Uni{1}_B$, which transforms the phases of all quarks equally and hence does not change $V_\text{CKM}$.

We can adapt a similar discussion to the lepton sector, but all complex phases in the charged current can be eliminated because of the zero neutrino masses.
Such elimination happens by taking $V_{\nu L} = V_{eL}$; see \cref{eq:CKM}.
This is possible because the mass matrix of neutrinos is zero and hence $V_{\nu L}$ can take arbitrary value.
Therefore, there are three U(1) symmetries, $V'_{\nu L} = V'_{eL} = \diag(e^{i\theta_e}, e^{i\theta_\mu}, e^{i\theta_\tau})$, that keep the mass matrix $M_e$ unchanged.
We write these U(1) transformations as $\Uni{1}_{L_i} (i=1,2,3)$.

Other operators that can violate CP are topological terms $\theta F\tilde{F}$.
There seem to be three topological terms in the SM corresponding to the three gauge symmetries.
However, since $\pi_3(\Uni{1})=0$, the $\Uni{1}_Y$ topological term $\theta_Y B\tilde{B}$ does not give any physical effect.\footnote{When the spacetime is compactified on $S^2 \times S^2$, and if there is $\Uni{1}_Y$ magnetic monopole (possibly from $\SU{5}$ breaking), $\theta_Y$ can become physical. We do not consider physics in the presence of magnetic monopoles in this paper.}
The theta factor of $\SU{3}_c$ topological term can generally be eliminated by chiral U(1) transformations of quarks, but $\Uni{1}_B$, which is vector-like transformation, cannot cause such elimination.
This unremovable CP-violating phase $\theta_\text{QCD}$ is well known as the strong CP phase.
Finally, $\SU{2}_W$ topological term can be removed by $\Uni{1}_B$ or $\Uni{1}_{L_i}$.
This is because $\SU{2}_W$ gauge interaction operates only on left-handed particles ($Q_i$ and $L_i$) and their anti-particles, and therefore even vector-like U(1) transformation can shift the theta factor of the topological term.
If we want to fix the $\SU{2}_W$ topological term to be zero, we have only three U(1) symmetries:
$\Uni{1}_{B-L}$, $\Uni{1}_{L_1-L_2}$, and $\Uni{1}_{L_2-L_3}$.
Practically, however, the effect of $\SU{2}_W$ instantons is highly suppressed by a factor $\exp(-8\pi^2/g_W^2)\sim 10^{-80}$ due to the small coupling constant $g_W$. If one ignores the $\SU{2}_W$ topological term, one can consider four $\Uni{1}$ symmetries such as $\Uni{1}_{B}$, $\Uni{1}_{L_1}$, $\Uni{1}_{L_2}$, and $\Uni{1}_{L_3}$ as rephasing transformations. Even in such a case, we confirmed that the classification of operators remains the same at least up to dimension 8. The lowest dimension operators where this distinction becomes important are dimension 9. For example, the dimension-nine operator $uuu e_1 L_2 L_3$ conserves $\Uni{1}_{B-L}$ and $\Uni{1}_{L_i-L_j}$, but violates $\Uni{1}_{B}$ and $\Uni{1}_{L_i}$.

\subsection{Definition of ``CP violation by a single operator''}
\label{sec:def_CPV}

In this paper, we use three different terms for operators: {\it CP-even}, {\it CP-odd}, and {\it CP-violating} operators.
The first two operators are defined as follows.
\begin{thm}
    An operator that is invariant under a CP transformation is a CP-even operator.
    An operator whose sign is reversed by a CP transformation is a CP-odd operator.
\end{thm}

A general operator $\mathcal{O}$ is either CP-even, CP-odd, or a linear combination of them.
All operators in the Lagrangian are gauge singlets.
Hence, they are transformed by $\CC_S \PP$ under CP transformation.
Since $(\CC_S \PP)^2 = 1$, the CP transformation law of operators is classified into three ways:\footnote{Here we assume that the operator $\mathcal{O}$ consists only of field operators and does not include its coefficient. Therefore, $\mathcal{O}$ does not have to be Hermitian by itself.}
\begin{itemize}
    \item[] \makebox[60pt]{Type I:} Not self-conjugate
        \begin{equation*}
            \mathcal{O} \xrightarrow{\ \ \CP \ \ } \mathcal{O}^{CP} \neq \pm\mathcal{O},
        \end{equation*}
        
    \item[] \makebox[60pt]{Type I\!I:} Self-conjugate
        \begin{equation*}
            \mathcal{O} \xrightarrow{\ \ \CP \ \ } \mathcal{O},
        \end{equation*}
    
    \item[] \makebox[60pt]{Type I\!I\!I:} Self-conjugate with a sign flip
    \begin{equation*}
        \mathcal{O} \xrightarrow{\ \ \CP \ \ } -\mathcal{O}.
    \end{equation*}
\end{itemize}

From this classification, we find two types of CP-even operators: 
the sum of a type-I operator and its conjugate, $\mathcal{O}+\mathcal{O}^{CP}$, and the type-I\!I operators themselves.
Similarly, there are two types of CP-odd operators:
the difference between a type-I operator and its conjugate, $\mathcal{O}-\mathcal{O}^{CP}$, and the type-I\!I\!I operators themselves.

Not all CP-odd operators cause CP violation because the above definition ignores the coefficients of operators.
For example, a type-I operator $\mathcal{O}$ has its complex coefficient $\lambda$ in the Lagrangian $\mathcal{L}$.
Because the Lagrangian has to be Hermitian, $\mathcal{L}$ has another term $\lambda^*\mathcal{O}^\dag$.
Because $\mathcal{O}^{CP}$ should be proportional to $\mathcal{O}^\dag$, we expect that $\mathcal{O}^{CP}=e^{i\phi}\mathcal{O}^\dag$.
In this case, these terms in $\mathcal{L}$ are written as
\begin{align}
    \lambda\mathcal{O} + \lambda^*\mathcal{O}^\dag
    &= \lambda\mathcal{O} + \lambda^*e^{-i\phi} \mathcal{O}^{CP} \notag \\
    &= e^{-i\phi/2}\qty(\lambda e^{i\phi/2}\mathcal{O} + \lambda^*e^{-i\phi/2} \mathcal{O}^{CP}) \notag \\
    &= e^{-i\phi/2}\qty[\Re(\lambda e^{i\phi/2})\qty(\mathcal{O}+\mathcal{O}^{CP}) +i\Im(\lambda e^{i\phi/2})\qty(\mathcal{O}-\mathcal{O}^{CP})]\, .
\end{align}
The second term seems to violate CP. However, sometimes the rephasing of fermions in $\mathcal{O}$ can remove the phase of $\lambda e^{i\phi/2}$, which is called the CP phase.
In such a case, we can take $\Im(\lambda e^{i\phi/2})=0$, and there is no CP violation by the CP-odd operator $\mathcal{O}-\mathcal{O}^{CP}$.
Therefore, we provide another definition for operators that cause CP violations.

\begin{thm}
    If a CP phase of an operator $\mathcal{O}$ cannot be removed by rephasing performed while preserving the CP phases in the SM Lagrangian, the CP-odd operator corresponding to $\mathcal{O}$ is a CP-violating operator.
\end{thm}

In the following part of this paper, we ignore $\SU{2}_W$ instanton effect.
In this case, as we discussed in \cref{sec:CPV_SM}, there are four U(1) symmetries that do not change $\delta_\text{CKM}$ and $\theta_\text{QCD}$.
If a CP-odd operator is not invariant under at least one of four U(1) transformations, its CP phase can be removed by the U(1) transformation. 
Therefore, CP-violating operators have to be invariant under all four U(1) transformations.

Note that CP-violating operators are defined as operators that violate CP when {\it only one} of them is added to the SM Lagrangian.
In general, however, when two or more CP-odd but not CP-violating operators are added to the SM  Lagrangian, they can cause CP violation.
This is because rephasing does not necessarily remove all of their CP phases simultaneously.
For example, suppose that there are $N$ CP-violating operators and $M$ CP-odd but not CP-violating operators, and that the latter contain at least one operator that is not invariant under each of the four U(1) transformations.
Then, we have $N+M-4$ independent CP phases. Since we can perform this counting once we have a complete list of CP-odd and CP-violating operators, we do not discuss the case with multiple operators any further in this paper.

\subsection{Implementation of CP into the Hilbert Series}
\label{subsec:even_odd}

The Hilbert series method is the way to systematically enumerate independent invariants under some group(s).
The previous research \cite{henning2017operator} presented a formula to count EFT operators modulo EOM and IBP redundancies.
This formula needs building blocks called ``spurions" to be assigned corresponding characters, which are traces of representation matrices of considered group(s).

For the Hilbert series for the SMEFT, spurions labeled by their field name consist of 
\begin{align}
    \phi\in\{H, Q, u, d, L, e, B_L, W_L, G_L\}\, ,
\end{align}
their CP partners 
\begin{align}
\phi^\dag\in 
\{H^\dag, Q^\dag, u^\dag, d^\dag, L^\dag, e^\dag, B_R, W_R, G_R\}\, ,
\end{align}
and the covariant derivative $\mathcal{D}.$
We can assign them appropriate characters of Lorentz and gauge groups.
In our case, where we want to list SMEFT operators based on the CP property, we need to construct the Hilbert series for invariants under Lorentz, gauge, and CP transformations as will be shown later.
Therefore, building blocks have to be in some representation of the group $(\SO{4}\times\SU{3}_c\times\SU{2}_W\times\Uni{1}_Y)\rtimes\Gamma_\CP$.
However, none of $\phi$ and $\phi^\dag$ has a definite representation under this group because they are transformed to each other by CP.
Therefore, we use the direct sum $\check{\phi}\equiv\phi\oplus\phi^\dag$ as building blocks.
Their representations under Lorentz and gauge groups are listed in \cref{tab:fields}.
Note that, when we use $\check{\phi}$ as building blocks, a CP-even operator $\mathcal{O}+\mathcal{O}^{CP}$ and a CP-odd operator $\mathcal{O}-\mathcal{O}^{CP}$ are written identically as $\check{\mathcal{O}}$.
For example, the dimension-five operator $\check{H}^2\check{L}^2$ means the direct sum $H^2L^2\oplus H^{\dag 2}L^{\dag 2}$, which includes the CP-even operator $H^2L^2 + H^{\dag 2}L^{\dag 2}$ and the CP-odd operator $H^2L^2 - H^{\dag 2}L^{\dag 2}$.

\begin{table}[t]
    \centering
    \[
    \renewcommand{\arraystretch}{1.3}
    \begin{array}{c|c}
         & (\SU{2}_l \times \SU{2}_r) \times (\SU{3}_c \times \SU{2}_W \times \Uni{1}_Y) \\ \hline
        \check{H} & \left[(\mathbf{1}, \mathbf{1}) \times (\mathbf{1},\mathbf{2})_{1/2}\right] \oplus \left[(\mathbf{1}, \mathbf{1}) \times (\mathbf{1},\mathbf{2})_{-1/2}\right] \\
        \check{Q} & \left[(\mathbf{2}, \mathbf{1}) \times (\mathbf{3},\mathbf{2})_{1/6}\right] \oplus  \left[(\mathbf{1}, \mathbf{2}) \times (\bar{\mathbf{3}},\mathbf{2})_{-1/6}\right] \\
        \check{u} & \left[(\mathbf{2}, \mathbf{1}) \times (\bar{\mathbf{3}},\mathbf{1})_{-2/3}\right] \oplus \left[(\mathbf{1}, \mathbf{2}) \times (\mathbf{3},\mathbf{1})_{2/3}\right] \\
        \check{d} & \left[(\mathbf{2}, \mathbf{1}) \times (\bar{\mathbf{3}},\mathbf{1})_{1/3}\right] \oplus \left[(\mathbf{1}, \mathbf{2}) \times (\mathbf{3},\mathbf{1})_{-1/3}\right] \\
        \check{L} & \left[(\mathbf{2}, \mathbf{1}) \times (\mathbf{1},\mathbf{2})_{-1/2}\right] \oplus \left[(\mathbf{1}, \mathbf{2}) \times (\mathbf{1},\mathbf{2})_{1/2}\right] \\
        \check{e} & \left[(\mathbf{2}, \mathbf{1}) \times (\mathbf{1},\mathbf{1})_{1}\right] \oplus \left[(\mathbf{1}, \mathbf{2}) \times (\mathbf{1},\mathbf{1})_{-1}\right] \\
        \check{G} & \left[(\mathbf{3}, \mathbf{1}) \times (\mathbf{8},\mathbf{1})_{0}\right] \oplus \left[(\mathbf{1}, \mathbf{3}) \times (\mathbf{8},\mathbf{1})_{0}\right] \\
        \check{W} & \left[(\mathbf{3}, \mathbf{1}) \times (\mathbf{1},\mathbf{3})_{0}\right] \oplus \left[(\mathbf{1}, \mathbf{3}) \times (\mathbf{1},\mathbf{3})_{0}\right] \\ 
        \check{B} & \left[(\mathbf{3}, \mathbf{1}) \times (\mathbf{1},\mathbf{1})_{0}\right] \oplus \left[(\mathbf{1}, \mathbf{3}) \times (\mathbf{1},\mathbf{1})_{0}\right] 
    \end{array}
    \]
    \renewcommand{\arraystretch}{1.0}
    \caption{The spurions for our Hilbert series method and their representations under Lorentz and gauge groups.}
    \label{tab:fields}
\end{table}

Let us see how to construct CP-even and odd Hilbert series for the SMEFT.
As derived in App.~C of \cite{henning2017operator} and \cite{graf20212}, 
they are expressed as linear combinations of two different Hilbert series,
\begin{align}
    H_{\text{even}} &= \frac{1}{2} \left(H_{+} + H_{-}\right), \\
    H_{\text{odd}} &= \frac{1}{2} \left(H_{+} - H_{-}\right). \label{eq:odd_HS}
\end{align}
The Hilbert series $H_+$ is composed of singlets under $\SO{4}\times\SU{3}\times\SU{2}\times\Uni{1}$, and $H_-$ consists of singlets under $(\SO{4}\times\SU{3}\times\SU{2}\times\Uni{1})\CP$.
They are given by\footnote{\Cref{eq:H_+,eq:H_-} need correction terms $\Delta H_\pm$ \cite{henning2017operator}, but they do not include operators with mass dimensions higher than four.
Since we have an interest in classifying only high-dimensional operators, we ignore $\Delta H_\pm$ terms in this paper.}
\begin{align}
    & H_{+}(\{\check{\phi_i}\}, \mathcal{D}) \notag\\
    &\ \ \equiv \int\mathrm{d}\mu_{\mathrm{O}_+(4)}(x)
        \int\mathrm{d}\mu_{\widetilde{\text{gauge}}_+}(y)
        \frac{1}{P_+(\mathcal{D},x)} \prod_i Z_{+}(\check{\phi_i},\mathcal{D},x,y)\, , \label{eq:H_+}\\
    & H_{-}(\{\check{\phi_i}\}, \mathcal{D}) \notag\\
    &\ \ \equiv \int\mathrm{d}\mu_{\mathrm{O}_-(4)}(\tilde{x})
        \int\mathrm{d}\mu_{\widetilde{\text{gauge}}_-}(\tilde{y})
        \frac{1}{P_-(\mathcal{D},\tilde{x})} \prod_i Z_{-}(\check{\phi_i},\mathcal{D},\tilde{x},\tilde{y})\, . \label{eq:H_-}
\end{align}
Here we used the notation for the Lorentz symmetry extended to the disconnected group by parity $\Gamma_\PP=\{1,\PP\}$,
\begin{align}
    \mathrm{O}(4) &= \SO{4} \rtimes \Gamma_\PP \notag \\ 
        &= \{\SO{4},\ \SO{4}\PP\} \notag \\
        &\equiv \{\mathrm{O}_+(4),\ \mathrm{O}_-(4)\}\, , 
\end{align}
and for the gauge symmetries extended by charge conjugation $\Gamma_\CC$,\footnote{For $\Gamma_{\CC_A}\cong\mathbb{Z}_4$, $\widetilde{\text{gauge}}$ seems to have four disconnected components.
Yet, because $\SU{2}$ includes the $\mathbb{Z}_2$ center, $\widetilde{\text{gauge}}$ consists of only two components as well as \cref{eq:gauge_tilde}. Therefore, the following discussion holds regardless of the choice of $\CC_S$ and $\CC_A$.}
\begin{align}
    \widetilde{\text{gauge}} 
        &\equiv (\SU{3} \times \SU{2} \times \Uni{1}) \rtimes \Gamma_\CC \notag \\
        &= \{\SU{3} \times \SU{2} \times \Uni{1},\ (\SU{3} \times \SU{2} \times \Uni{1})\CC\} \notag \\
        & \equiv \{\widetilde{\mathrm{SU}}_+(3) \times \widetilde{\mathrm{SU}}_+(2) \times \widetilde{\mathrm{U}}_+(1),\ \widetilde{\mathrm{SU}}_-(3) \times \widetilde{\mathrm{SU}}_-(2) \times \widetilde{\mathrm{U}}_-(1)\} \notag \\
        & \equiv \{\widetilde{\text{gauge}}_+,\ \widetilde{\text{gauge}}_-\}\, .
    \label{eq:gauge_tilde}
\end{align}

We summarize the details of \cref{eq:H_+,eq:H_-} below.

\begin{enumerate}
    \item
        The integrands $Z_{\pm}$ are calculated by
        \begin{align}
            &Z_+(\check{\phi_i},\mathcal{D},x,y) =
                \mathrm{exp}\left[
                    \sum_{n=1}^\infty (\pm 1)^{n+1} \frac{1}{n} \left(\frac{\check{\phi_i}}{\mathcal{D}^{\Delta_i}}\right)^n \tr(g_{i,+}^n)
                \right], \\
            &Z_-(\check{\phi_i},\mathcal{D},\tilde{x},\tilde{y}) =
                \mathrm{exp}\left[
                    \sum_{n=1}^\infty (\pm 1)^{n+1} \frac{1}{n} \left(\frac{\check{\phi_i}}{\mathcal{D}^{\Delta_i}}\right)^n \tr(g_{i,-}^n)
                \right].
        \end{align}
        The sign $\pm$ is plus for bosonic fields and minus for fermionic fields.
        Thanks to this, we can take the coefficient of $(\check{\phi_i}/\mathcal{D}^{\Delta_i})^n$ to be $n$th symmetric or anti-symmetric product of $\tr(g_i)$, {\it i.e.} $\mathrm{sym}^n[\tr(g_i)]$ or $\land^n[\tr(g_i)]$, reflecting the statistics of $\check{\phi_i}$.
        The exponent $\Delta_i$ is the mass dimension of $\check{\phi_i}$.
        
        The representation matrices $g_{i,+}(\mathcal{D},x,y)$ and $g_{i,-}(\mathcal{D},\tilde{x},\tilde{y})$ are defined as a tensor product of those for the spacetime and internal symmetry groups:
        \begin{align}
            &g_{i,+}(\mathcal{D},x,y) \equiv g^{\mathrm{O}_+(4)}_i(\mathcal{D},x) \otimes g^{\widetilde{\text{gauge}}_+}_i(y)\, , \\
            &g_{i,-}(\mathcal{D},\tilde{x},\tilde{y}) \equiv g^{\mathrm{O}_-(4)}_i(\mathcal{D},\tilde{x}) \otimes g^{\widetilde{\text{gauge}}_-}_i(\tilde{y})\, .
        \end{align}
        The variables $x\equiv(x_1,x_2)$ and $y\equiv (y_{c,1},y_{c,2},y_{W},y_Y)$ parametrize maximal tori of the groups $\mathrm{O}_+(4)=\SO{4}$ and $\widetilde{\text{gauge}}_+=\SU{3}_c\times \SU{2}_W \times \Uni{1}_Y$ respectively, and
        $\tilde{x}\equiv x_1$, $\tilde{y}\equiv (y_{c,1},y_W)$.
        
        On the $H_+$ branch, $\tr(g_{i,+}^n)$ is written in terms of the character $\chi_{i,+}=\tr(g_{i,+})$ as
        \begin{align}
            &\tr(g_{i,+}^n(\mathcal{D},x,y)) \notag \\
                &\quad = \chi_{i,+}(\mathcal{D}^n,x^n,y^n) \notag \\
                &\quad = \chi^{\mathrm{O}_+(4)}_{i}(\mathcal{D}^n,x^n) \;         \chi^{\widetilde{\text{gauge}}_+}_i(y^n) \notag \\
                &\quad = \chi^{\SO{4}}_{i}(\mathcal{D}^n,x^n) \; \chi^{\SU{3}}_i(y_c^n) \; \chi^{\SU{2}}_i(y_W^n) \; \chi^{\Uni{1}}_i(y_Y^n)\, . \label{eq:g_+}
        \end{align}
        Here we adopted an abbreviated notation $x^n=(x_1^n,x_2^n)$ as well as $y_c^n$.
        The formulae for the characters are listed in \cref{sec:char}.
        
        On the contrary, we cannot apply the same method for $H_-$ branch: $\tr(g_{i,-}^n(\mathcal{D},\tilde{x},\tilde{y}))\neq\chi_{i,-}(\mathcal{D}^n,\tilde{x}^n,\tilde{y}^n)$.
        Instead, we have
        \begin{equation}
            \tr(g_{i,-}^n(\mathcal{D},\tilde{x},\tilde{y})) = \left\{
                \begin{aligned}
                    & 0 && (n: \text{odd}) \\
                    & \pm \chi_{i,+}(\mathcal{D}^n,\bar{x}^n,\bar{y}^n) && (n: \text{even})
                \end{aligned}
            \right. \ ,
            \label{eq:g_-}
        \end{equation}
        where $\bar{x}\equiv(x_1,1)$ and $\bar{y}\equiv(y_{c,1},1,y_W,1)$; see App.~C of \cite{henning2017operator} and \cite{graf20212}.
        The zero value for odd $n$ comes from the fact that none of the field contents is invariant under CP transformation.
        The sign $\pm$ is plus for the choice of $\CC_S\PP$, and minus for $\CC_A\PP$ corresponding to the sign of squared CP transformation.
    
    \item 
        The factor $P_{\pm}$ removes IBP redundancies and is written as
        \begin{align}
            &P_+(\mathcal{D},x) = \frac{1}{(1-\mathcal{D}x_1)(1-\mathcal{D}x_1^{-1})(1-\mathcal{D}x_2)(1-\mathcal{D}x_2^{-1})}\, , \label{eq:P_+}\\
            &P_-(\mathcal{D},\tilde{x}) = \frac{1-\mathcal{D}^2}{(1-\mathcal{D}x_1)(1-\mathcal{D}x_1^{-1})}\, .
        \end{align}

    \item
        The Haar measures for the $H_+$ branch are
        \begin{align}
            \mathrm{d}\mu_{\mathrm{O}_+(4)}(x_1) &= \mathrm{d}\mu_{\SO{4}}(x_1, x_2)\, ,\\
            \mathrm{d}\mu_{\widetilde{\text{gauge}}_+}(\tilde{y}) &= \mathrm{d}\mu_{\SU{3}}(y_{c,1},y_{c,2}) \, \mathrm{d}\mu_{\SU{2}}(y_W) \, \mathrm{d}\mu_{\Uni{1}}(y_Y)\, .
        \end{align}
        For the $H_-$ branch,
        \begin{align}
            \mathrm{d}\mu_{\mathrm{O}_-(4)}(x_1) &= \mathrm{d}\mu_{\mathrm{Sp}(2)}(x_1)\, , \label{eq:Haar_O-}\\
            \mathrm{d}\mu_{\widetilde{\text{gauge}}_-}(\tilde{y}) &= \mathrm{d}\mu_{\widetilde{\mathrm{SU}}_-(3)}(y_{c,1}) \, \mathrm{d}\mu_{\widetilde{\mathrm{SU}}_-(2)}(y_W)\, , \\
            \mathrm{d}\mu_{\widetilde{\mathrm{SU}}_-(3)}(y_{c,1}) &= \mathrm{d}\mu_{\mathrm{Sp}(2)}(y_{c,1})\, , \label{eq:Haar_SU3-}\\
            \mathrm{d}\mu_{\widetilde{\mathrm{SU}}_-(2)}(y_W) &=  \left\{
            \begin{aligned}
                &\mathrm{d}\mu_{\SU{2}}(iy_W) && \text{for} \ \ \CC_S\, , \\
                &\mathrm{d}\mu_{\SU{2}}(y_W) && \text{for} \ \ \CC_A\, .
            \end{aligned}
            \right.\label{eq:Haar_SU2-}
        \end{align}
        \Cref{eq:Haar_O-,eq:Haar_SU3-} are derived with the ``folding" technique in \cite{henning2017operator} and \cite{graf20212}, and \cref{eq:Haar_SU2-} is derived in \cite{arias2019discrete}.
        The formulae for the Haar measures are listed in \cref{sec:char}.
        
        For $\widetilde{\mathrm{SU}}_{-}(2)$, if one tries to remove the factor $i$ in \cref{eq:Haar_SU2-} for $\CC_S$ by changing variable $y_W\to y_W'=-iy_W$,
        then the sign of \cref{eq:g_-} will be reversed:
        $\mathrm{tr}(g_-^n) (p,x_1,y_{c,1},y_W) \to -\mathrm{tr}(g_-^n) (p,x_1,y_{c,1},y_W')$.
        Therefore, the same result can be obtained whichever you choose $\CC_S$ or $\CC_A$. 
        This result is expected from \cref{eq:equivalence}.
        Yet it is clear that the choice of $\CC_A$ is more ``natural'' in this formulation.

\end{enumerate}

CP-violating operators defined in \cref{sec:def_CPV} are obtained by a simple modification; just add four U(1) symmetries, $\Uni{1}_{B}$ and $\Uni{1}_{L_i}$ ($i=1,2,3$) to the Hilbert series $H_\pm$.
All you need is two procedures.
First, assign these U(1) charges to the character of each building block.
Second, integrate with four U(1) Haar measures.

\section{Result}
\label{sec:result}
In this section, we will summarize some results of counting.
These results can be reproduced by an auxiliary \FORM file that computes the full Hilbert series for the SMEFT.
This file is based on the code called ECO (Effective Counting of Operators) provided in \cite{marinissen202083106786}.

\subsection{List of operators at dimension eight}

\begin{table}\footnotesize
\renewcommand\arraystretch{1.4}
\vspace{-5\baselineskip}
\[
\hspace{-2cm}
\begin{array}{cc|crr|crrr}
\text{Class} & & \multicolumn{3}{c|}{\text{CP-even}} & \multicolumn{4}{c}{\text{CP-odd}}\\
 & & N_f & 1 & 3 & N_f & 1 & 3 & 3'\\
\hline
X^4 & & 26 & 26 & 26 & 17 & 17 & 17 & 17\\
X^3 H^2 & & 3 & 3 & 3 & 3 & 3 & 3 & 3\\
X^2 H^4 & & 5 & 5 & 5 & 5 & 5 & 5 & 5\\
H^8 & & 1 & 1 & 1 & 0 & 0 & 0 & 0\\
X^2H \psi^2 & & 48N_f^2 & 48 & 432 & 48N_f^2 & 48 & 432 & 372\\
XH^3 \psi^2 & & 11N_f^2 & 11 & 99 & 11N_f^2 & 11 & 99 & 81\\
H^5 \psi^2 & & 3N_f^2 & 3 & 27 & 3N_f^2 & 3 & 27 & 21\\
\multirow{2}{*}{$X \psi^4$} & \hspace{-10pt}(B) & 2N_f^2(40N_f^2 - 1) & 78 & 6462 & 2N_f^2(40N_f^2 - 1) & 78 & 6462 & 4344\\
 & \hspace{-10pt}(\slashed{B}) & N_f^3(21N_f +1) & 22 & 1728 & N_f^3(21N_f +1) & 22 & 1728 & 0\\
\multirow{2}{*}{$H^2\psi^4$} & \hspace{-10pt}(B) & \frac{1}{2}N_f(67N_f^3 + N_f^2 + 37N_f + 1) & 53 & 2895 & \frac{1}{2}N_f(67N_f^3 + N_f^2 - 23N_f - 1) & 22 & 2622 & 1566 \\
& \hspace{-10pt}(\slashed{B}) & \frac{1}{6}N_f^2(43N_f^2 - 9N_f + 2) & 6 & 543 & \frac{1}{6}N_f^2(43N_f^2 - 9N_f + 2) & 6 & 543 & 0\\
\hline
X^2\psi^2\scD & & \frac{1}{2}N_f(57N_f + 23) & 40 & 291 & \frac{1}{2}N_f(57N_f - 23) & 17 & 222 & 183\\
XH^2\psi^2\scD & & 46N_f^2 & 46 & 414 & 46N_f^2 & 46 & 414 & 330\\
H^4\psi^2\scD & & \frac{1}{2}N_f(13N_f + 7) & 10 & 69 & \frac{1}{2}N_f(13N_f - 7) & 3 & 48 & 33\\
\multirow{2}{*}{$H\psi^4\scD$} & \hspace{-10pt}(B) & \frac{1}{2}N_f^3(135N_f-1) & 67 & 5454 & \frac{1}{2}N_f^3(135N_f-1) & 67 & 5454 & 3387\\
& \hspace{-10pt}(\slashed{B}) & \frac{1}{2}N_f^3(29N_f+3) & 16 & 1215 & \frac{1}{2}N_f^3(29N_f+3) & 16 & 1215 & 0\\
\hline
X^2H^2\scD^2 & & 11 & 11 & 11 & 7 & 7 & 7 & 7\\
XH^4\scD^2 & & 3 & 3 & 3 & 3 & 3 & 3 & 3\\
H^6\scD^2 & & 2 & 2 & 2 & 0 & 0 & 0 & 0\\
XH\psi^2\scD^2 & & 24N_f^2 & 24 & 216 & 24N_f^2 & 24 & 216 & 180\\
H^3\psi^2\scD^2 & & 18N_f^2 & 18 & 162 & 18N_f^2 & 18 & 162 & 126\\
\multirow{2}{*}{$\psi^4\scD^2$} & \hspace{-10pt}(B) & \frac{1}{4}N_f^2(99N_f^2 + 89) & 47 & 2205 & \frac{1}{4}N_f^2(99N_f^2 - 67) & 8 & 1854 & 1086\\
& \hspace{-10pt}(\slashed{B}) & \frac{1}{2}N_f^3(11N_f - 1) & 5 & 432 & \frac{1}{2}N_f^3(11N_f - 1) & 5 & 432 & 0\\
\hline
H^2\psi^2\scD^3 & & N_f(8N_f + 7) & 15 & 93 & N_f(8N_f - 7) & 1 & 51 & 33\\
\hline
H^4\scD^4 & & 3 & 3 & 3 & 0 & 0 & 0 & 0\\
\hline
\hline
\multirow{2}{*}{Total} & \hspace{-10pt}(B) & \frac{823}{4}N_f^4 + \frac{929}{4} N_f^2 + 22N_f + 54 & 514 & 18873 & \frac{823}{4}N_f^4 + \frac{649}{4} N_f^2 - 22N_f + 35 & 381 & 18098 & 11777 \\
& \hspace{-10pt}(\slashed{B}) & \frac{289}{6}N_f^4 + \frac{1}{2}N_f^3 + \frac{1}{3} N_f^2  & 49 & 3918 & \frac{289}{6}N_f^4 + \frac{1}{2}N_f^3 + \frac{1}{3} N_f^2  & 49 & 3918 & 0
\end{array}
\]
\caption{\label{tbl:dim8} The number of dimension-eight operators for arbitrary $N_f$ as well as for $N_f=1$ and $N_f=3$. 
The horizontal lines separate the operators according to the number of derivatives involved.
Operators with four fermions are further distinguished by whether they preserve the baryon number ($B$) or violate the baryon number ($\slashed{B}$).
The last column with $N_f=3'$ indicates the number of CP-violating operators defined in \cref{sec:def_CPV}.}
\end{table}

First, we categorize the dimension-eight operators into several classes in \cref{tbl:dim8}.
Here we use $H$ for Higgs $\check{H}$, $\psi$ for fermions $\check{Q},\check{u},\check{d},\check{L},\check{e}$, and $X$ for field strengths $\check{B},\check{W},\check{G}$ as well as \cite{lehman2016low}.
In total, we find 430 CP-odd operators for $N_f=1$, and 22016 (11777) CP-odd (-violating) operators for $N_f=3$.

We find that the existing classification \cite{remmen2019consistency}, which counts dimension-eight CP-even and CP-odd operators involving only bosonic fields, differs from our result in the classification of operators $\check{H}^2 \scD^2 \check{W}^2$. 
The authors claim that the operator $\check{H}^2 \scD^2 \check{W}^2$ involves 3 CP-even operators and 3 CP-odd operators.
In contrast, our result with the Hilbert series shows that there are 4 CP-even and 2 CP-odd operators.
To understand the discrepancy, let us consider the operator $i\epsilon^{IJK}(\scD^\mu H^\dag \sigma^I \scD^\nu H)(W^J_{\mu\rho}\widetilde{W}_\nu^{K\rho}+\widetilde{W}^J_{\mu\rho}W_\nu^{K\rho})$.
This is classified into CP-odd operators in \cite{remmen2019consistency}, but it is actually a CP-even operator.
This can be shown using either $\CC_S$ or $\CC_A$ as charge conjugation, and we will use $\CC_A$ here.
The example of the matrix $C_A$ in the asymmetric charge conjugation $\CC_A$ for $\SU{2}$ has already appeared in \cref{eq:H_Ca}, $C_A = i\sigma^2$.
Following to \cref{eq:CC_adj}, $W \equiv W^I \sigma^I / 2$ transforms as
\begin{align}
    W \xrightarrow{\ \CC_A \ } & -(i\sigma^2)^* W^T (i\sigma^2)^T \notag \\
        & = (-W^T)^* \notag \\
        & = -W\, .
    \label{eq:W_Ca}
\end{align}
Here, we have used $\sigma^2 \sigma^I \sigma^2 = -\sigma^{I*}$.
From \cref{eq:W_Ca}, we find 
\begin{equation}
    W^I \xrightarrow{\ \CC_A \ } - W^I
    \quad \text{for $I = 1,2,3$}\, .
\end{equation}
Considering that parity $\PP$ flips the sign of $W\widetilde{W}$, we find
\begin{align}
& i\epsilon^{IJK}[\mathcal{D}^\mu H^{\dag} \sigma^{I} \mathcal{D}^\nu H](W^J_{\mu\rho}\widetilde{W}_{\nu}^{K\rho}+\widetilde{W}^J_{\mu\rho}W_{\nu}^{K\rho}) \notag \\
& \xrightarrow{\ \CC_A\PP \ }
  i\epsilon^{IJK}[\mathcal{D}^\mu H (i\sigma^2)^T \sigma^{I} (i\sigma^2)^* \mathcal{D}^\nu H^\dag][-(-W^J_{\mu\rho})(-\widetilde{W}_{\nu}^{K\rho})-(-\widetilde{W}^J_{\mu\rho})(-W_{\nu}^{K\rho})] \notag \\
& \quad = - i\epsilon^{IJK}[\mathcal{D}^\nu H^\dag \sigma^{2\dag} (\sigma^{I})^T \sigma^2 \mathcal{D}^\mu H][W^J_{\mu\rho}\widetilde{W}_{\nu}^{K\rho}+\widetilde{W}^J_{\mu\rho}W_{\nu}^{K\rho}] \notag \\
& \quad = + i\epsilon^{IJK}[\mathcal{D}^\mu H^\dag \sigma^{I} \mathcal{D}^\nu H][W^J_{\mu\rho}\widetilde{W}_{\nu}^{K\rho}+\widetilde{W}^J_{\mu\rho}W_{\nu}^{K\rho}]\, .
\end{align}
Hence, this operator is CP even, and our Hilbert series method identifies it correctly.

\subsection{The number of operators up to dimension 14}

\begin{table}[t]\footnotesize
    \hspace{-2.0cm}
    \begin{tabular}{c|l|l l l l l l l l l l}
        $N_f$ & Dimension$=$ & 5 & 6 & 7 & 8 & 9 & 10 & 11 & 12 & 13 & 14 \\ \hline
        1 & CP-odd & 1 & 27 & 15 & 430 & 280 & 7414 & 6981 & 129253 & 128689 & 2300238 \\
        1 & CP-violating & 0 & 23 & 0 & 381 & 0 & 6242 & 0 & 103268 & 0 & 1743183 \\ \hline
        3 & CP-odd & 6 & 1422 & 771 & 22016 & 45228 & 1042942 & 1736133 & 37761366 & 87686796 & 1397416310 \\
        3 & CP-violating & 0 & 705 & 0 & 11777 & 0 (60) & 437331 & 0 (5757) & 13891774 & 0 (528689) & 454073102 \\ \hline
        \multicolumn{2}{l|}{Time (in seconds)} & 0.01 & 0.03 & 0.05 & 0.12 & 0.25 & 0.59 & 1.20 & 3.18 & 6.11 & 14.25
    \end{tabular}
    \caption{The numbers of CP-odd and CP-violating operators for $N_f=1, 3$, as well as the computing time for $N_f=3$ CP-odd operators on the Apple M1 max with ten cores.
    The numbers in parentheses are those of CP-violating operators when the $\SU{2}_W$ instanton effect is taken into account.}
    \label{tbl:time}
\end{table}

\begin{figure}[t]
\centering
\includegraphics[width=13cm]{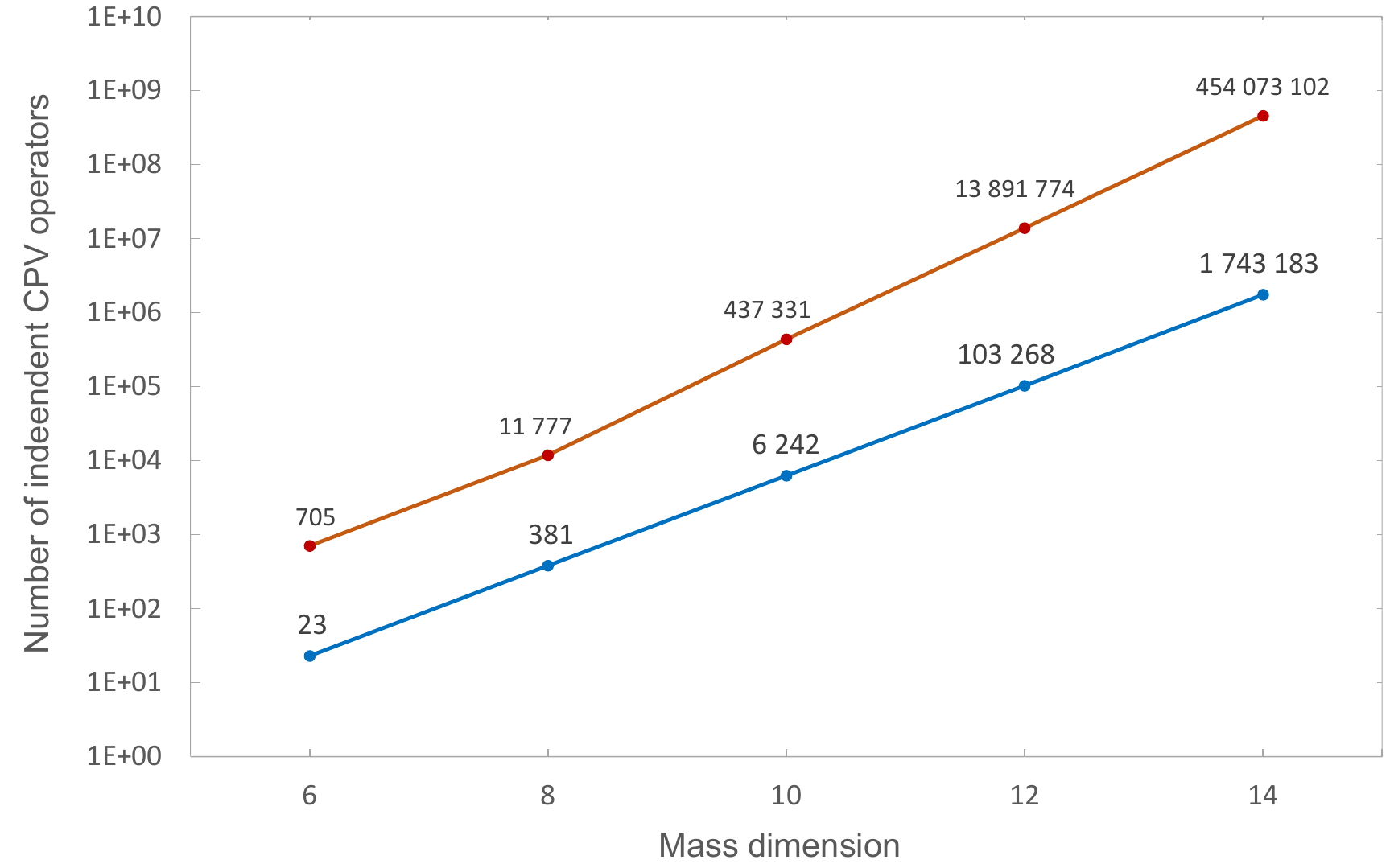}
\caption{The number of independent CP-violating operators in the SMEFT up to mass dimension 14. Points joined by the lower solid line are for $N_f=1$; those joined by the upper solid line are for $N_f=3$. All operators of odd dimensions can have their phases removed by rephasing up to this dimension, and hence they are not shown.}
\label{fig:growth}
\end{figure}

We summarize the numbers of CP-odd and CP-violating operators for $N_f=1,3$ in \cref{tbl:time}.
This table shows that there are 1422 CP-odd and 705 CP-violating dimension-six operators for $N_f=3$.
The 1422 CP-odd operators are made up of 273 operators violating the baryon number and 1149 operators conserving the baryon number, which is consistent with the counting by \cite{alonso2014renormalization}.
The 705 CP-violating operators consist of 6 bosonic operators and 699 operators including fermions, and this is consistent with the counting by \cite{bonnefoy2021beyond}.

We also visualized the growth of the number of CP-violating operators up to dimension 14 in \cref{fig:growth}.

\section{Conclusion and Discussion}
\label{sec:conclusion}

In this paper, we have shown how to classify the SMEFT operators based on CP property by means of the Hilbert series techniques.
We successfully reproduced the same enumerations as those by \cite{alonso2014renormalization,bonnefoy2021beyond} for dimension-six operators and pointed out a misidentification by \cite{remmen2019consistency} for dimension-eight operators.
Our \FORM code can output these results in a few seconds and can list higher-dimensional operators quickly.
Our method can be easily applied to other EFT theories besides the SMEFT, such as QCD EFT or the SMEFT with gravity.

As is mentioned at the end of \cref{sec:def_CPV}, our method correctly lists operators that violate CP that cannot be removed by rephasing when only one of them is added to SMEFT. In the presence of multiple operators, CP-conserving operators on our list can conspire together to violate CP as there is less freedom for rephasing. The classification of such a possibility is beyond the scope of this paper.

\section*{Note Added}

While we were finalizing this paper, the paper \cite{Sun:2022aag} appeared. They consider Higgs Effective Field Theory (HEFT), not SMEFT, with CP transformation. They introduce right-handed neutrinos and another  $\SU{2}_R$ gauge to allow for separate C and P transformations, which we do not need in our formulation. In addition, they do not consider removing CP-violating phases by rephasing either. Some details of the implementations of CP also are different, and the consistency between theirs and ours will be studied elsewhere, given that SMEFT and HEFT operators cannot be compared directly.

\section*{Acknowledgment} 
HM thanks Brian Henning, Xiaochuan Lu, and Tom Melia for a long-term fruitful collaboration on this subject.
The works of DK and HM were supported by the Institute for AI and Beyond of the University of Tokyo.
The work of HM was also supported by the Director, Office of Science, Office of High Energy Physics of the U.S. Department of Energy under the Contract No. DE-AC02-05CH11231, by the NSF grant PHY-1915314, by the Binational Science Foundation (grant No. 2018140), by the JSPS Grant-in-Aid for Scientific Research JP20K03942, MEXT Grant-in-Aid for Transformative Research Areas (A) JP20H05850, JP20A203, by WPI, MEXT, Japan, and Hamamatsu Photonics, K.K. 
The work of RO was supported by Forefront Physics and Mathematics Program to Drive Transformation (FoPM), a World-leading Innovative Graduate Study (WINGS) Program, the University of Tokyo.

\appendix
\section{Unitarity of charge conjugation}
\label{sec:unitary_CC}

It is sometimes misunderstood that the charge conjugation is anti-unitary given that it involves a complex conjugation. In this section, we clarify why the apparent complex conjugation is consistent with charge conjugation being unitary. We demonstrate how to define unitary charge conjugation $\CC$ that is consistent with \cref{eq:CC_Phi2,eq:CC_Phidag3}. Anti-unitary operations appear only when they involve time reversal.

As in \cref{subsec:CC_complex}, let $\Phi=(\phi_1,\ \cdots,\ \phi_N)^T$ be a scalar quantum field operator in the representation ${\bf N}$.
We write them and their Hermitian conjugates in terms of the annihilation operators $a_i$ and $b_i$ as
\begin{align}
  \Phi(x) &= \mqty(\phi_1(x) \\ \vdots \\ \phi_N(x))
    = \int \frac{\mathrm{d}^3p}{(2\pi)^3 2E} \left[
        \mqty(a_1({\bf p}) \\ \vdots \\ a_N({\bf p})) e^{-ip\cdot x}
      + \mqty( b_1^\dag({\bf p}) \\ \vdots \\ b_N^\dag({\bf p})) e^{+ip\cdot x}
      \right]\, ,\\
  \Phi^\dag(x) &= \mqty(\phi_1^*(x) \\ \vdots \\ \phi_N^*(x))
    = \int \frac{\mathrm{d}^3p}{(2\pi)^3 2E} \left[
        \mqty(a_1^\dag({\bf p}) \\ \vdots \\ a_N^\dag({\bf p})) e^{+ip\cdot x}
      + \mqty(b_1({\bf p}) \\ \vdots \\  b_N({\bf p})) e^{-ip\cdot x}
      \right]\, .
\end{align}
Here, we assumed that $\Phi$ is a scalar field for simplicity, but
the following discussion can be applied to a general field.

By comparing the coefficients of $e^{\pm ip\cdot x}$ in \cref{eq:CC_Phi2,eq:CC_Phidag3}, we find the transformation laws of creation and annihilation operators:
\begin{align}
  \mqty(a_1 \\ \vdots \\ a_N)\,
    &\xrightarrow{\ \ \CC\ \ }\,
    C \mqty(b_1 \\ \vdots \\ b_N)\, , \label{eq:CC_a}\\
  \mqty(b_1^\dag \\ \vdots \\ b_N^\dag)\,
    &\xrightarrow{\ \ \CC\ \ }\,
    C \mqty(a_1^\dag \\ \vdots \\ a_N^\dag)\, \label{eq:CC_bdag}.
\end{align}
These give the definition of the unitary charge conjugation operator.
Conversely, \cref{eq:CC_a,eq:CC_bdag} cannot be consistent if $\CC$ is antiunitary because antiunitary $\CC$ would transform $e^{\pm ip\cdot x}$ to $e^{\mp ip\cdot x}$.

\section{Characters and Haar measures}
\label{sec:char}
In \cref{tbl:character,tbl:Haar}, we summarize characters and Haar measures we used in \cref{subsec:even_odd}.
They are described by the maximal tori of the groups.
The redundancy of EOM can be removed by using the formulae of characters for the Lorentz group \cite{henning2017operator}.

For example, let us write down the characters for the left-handed quark doublet $\check{Q}$.
Because its representation of the group $(\SU{2}_l\times \SU{2}_r) \times (\SU{3}_c\times \SU{2}_W \times \Uni{1}_Y)$ is $\left[(\mathbf{2}, \mathbf{1}) \times (\mathbf{3},\mathbf{2})_{1/6}\right] \oplus  \left[(\mathbf{1}, \mathbf{2}) \times (\bar{\mathbf{3}},\mathbf{2})_{-1/6}\right]$, the character $\chi_+$ for $\check{Q}$ is written as
\begin{align}
    \chi_{\check{Q},+}(\mathcal{D},x,y)  
    &= \chi^{\text{Lorentz}}_{\text{LH spnior}}(\mathcal{D},x) \; \chi^{\SU{3}}_{\mathbf{3}}(y_c) \; \chi^{\SU{2}}_{\mathbf{2}}(y_W) \; \chi^{\Uni{1}}_{1/6}(y_Y)  \notag \\
    & + \chi^{\text{Lorentz}}_{\text{RH spnior}}(\mathcal{D},x) \; \chi^{\SU{3}}_{\bar{\mathbf{3}}}(y_c) \; \chi^{\SU{2}}_{\mathbf{2}}(y_W) \; \chi^{\Uni{1}}_{-1/6}(y_Y)\,.
\end{align}

\begin{table}[tbh]
    \centering
    \renewcommand{\arraystretch}{1.5}
    \[
    \hspace{-1cm}
    \begin{array}{l|c|c}
        \text{Group} & \text{Rep.} & \text{Character} \\ \hline
        \text{Lorentz} & \text{scalar} &  \mathcal{D}P_+(\mathcal{D},x_1,x_2)(1-\mathcal{D}^2)  \\
            & \text{LH spinor} & \mathcal{D}^{\frac{3}{2}}P_+(\mathcal{D},x_1,x_2)\left(x_1+\frac{1}{x_1}-\mathcal{D}\left(x_2+\frac{1}{x_2}\right)\right)  \\
            & \text{RH spnior} & \mathcal{D}^{\frac{3}{2}}P_+(\mathcal{D},x_1,x_2)\left(x_2+\frac{1}{x_2}-\mathcal{D}\left(x_1+\frac{1}{x_1}\right)\right) \\
            & \text{LH field strength} & \mathcal{D}^2P_+(\mathcal{D},x_1,x_2)\left(x_1^2+1+\frac{1}{x_1^2}-\mathcal{D}\left(x_1+\frac{1}{x_1}\right)\left(x_2+\frac{1}{x_2}\right)+\mathcal{D}^2\right) \\
            & \text{RH field strength} & \mathcal{D}^2P_+(\mathcal{D},x_1,x_2)\left(x_2^2+1+\frac{1}{x_2^2}-\mathcal{D}\left(x_1+\frac{1}{x_1}\right)\left(x_2+\frac{1}{x_2}\right)+\mathcal{D}^2\right) \\ \hline 
        \SU{3}_c & \mathbf{1} & 1 \\
            & \mathbf{3} & y_{c,1} + \frac{y_{c,2}}{y_{c,1}} + \frac{1}{y_{c,2}} \\
            & \overline{\mathbf{3}} & y_{c,2} + \frac{y_{c,1}}{y_{c,2}} + \frac{1}{y_{c,1}}  \\
            & \mathbf{8} & y_{c,1}y_{c,2} + \frac{y_{c,2}^2}{y_{c,1}} + \frac{y_{c,1}^2}{y_{c,2}} + 2 + \frac{y_{c,2}}{y_{c,1}^2} + \frac{y_{c,1}}{y_{c,2}^2} + \frac{1}{y_{c,1}y_{c,2}} \\ \hline 
        \SU{2}_W & \mathbf{1} & 1 \\
            & \mathbf{2} & y_W + \frac{1}{y_W} \\
            & \mathbf{3} & y_W^2 + 1 + \frac{1}{y_W^2} \\ \hline
        \Uni{1}_Y & Q & y_Y^Q  
    \end{array}
    \]
    \caption{List of characters used to calculate the Hilbert series. We use parameters of maximal tori $x_1$ and $x_2$ for $\SU{2}_l$ and $\SU{2}_r$ respectively, $(y_{c,1}, y_{c,2})$ for $\SU{3}_c$, $y_W$ for $\SU{2}_W$, and $y_Y$ for $\Uni{1}_Y$. The spurion $\mathcal{D}$ represents the covariant derivative. $P_+(\mathcal{D},x_1,x_2)$ is given by \cref{eq:P_+}.}
    \label{tbl:character}
\end{table}

\begin{table}[t]
    \[
    \begin{array}{l|c}
        \text{Group} & \text{Haar measure} \\ \hline
        \text{Lorentz} & \frac{\mathrm{d}x_1}{2\pi ix_1}\frac{\mathrm{d}x_1}{2\pi ix_1}(1-x_1^2)(1-x_2^2) \\
        \SU{3}_c & \frac{\mathrm{d}y_{c,1}}{2\pi iy_{c,1}}\frac{\mathrm{d}y_{c,2}}{2\pi iy_{c,2}}(1-y_{c,1}y_{c,2})\left(1-\frac{y_{c,1}^2}{y_{c,2}}\right)\left(1-\frac{y_{c,2}^2}{y_{c,1}}\right) \\
        \SU{2}_W & \frac{\mathrm{d}y_W}{2\pi iy_W}(1-y_W^2) \\
        \Uni{1}_Y & \frac{\mathrm{d}y_Y}{2\pi iy_Y}
    \end{array}
    \]
    
    \renewcommand{\arraystretch}{1}
    \caption{List of Haar measures used to calculate the Hilbert series. See the caption of \cref{tbl:character} for the definition of variables.}
    \label{tbl:Haar}
\end{table}

\clearpage

\bibliographystyle{utcaps_mod}
\bibliography{main}

\providecommand{\href}[2]{#2}\begingroup\raggedright\begin{thebibliography}{10}

\bibitem{Jenkins:2009dy}
E.~E. Jenkins and A.~V. Manohar, ``{\em {Algebraic Structure of Lepton and
  Quark Flavor Invariants and CP Violation}},''
  \href{http://dx.doi.org/10.1088/1126-6708/2009/10/094}{JHEP {\normalfont
  \bfseries 10} (2009)  094},
  \href{http://arxiv.org/abs/0907.4763}{{\normalfont \ttfamily
  arXiv:0907.4763}}.

\bibitem{Lehman:2015via}
L.~Lehman and A.~Martin, ``{\em {Hilbert Series for Constructing Lagrangians:
  expanding the phenomenologist's toolbox}},''
  \href{http://dx.doi.org/10.1103/PhysRevD.91.105014}{Phys. Rev. D {\normalfont
  \bfseries 91} (2015)  105014},
  \href{http://arxiv.org/abs/1503.07537}{{\normalfont \ttfamily
  arXiv:1503.07537}}.

\bibitem{Henning:2015daa}
B.~Henning, X.~Lu, T.~Melia, and H.~Murayama, ``{\em {Hilbert series and
  operator bases with derivatives in effective field theories}},''
  \href{http://dx.doi.org/10.1007/s00220-015-2518-2}{Commun. Math. Phys.
  {\normalfont \bfseries 347} (2016) no.~2, 363--388},
  \href{http://arxiv.org/abs/1507.07240}{{\normalfont \ttfamily
  arXiv:1507.07240}}.

\bibitem{lehman2016low}
L.~Lehman and A.~Martin, ``{\em {Low-derivative operators of the Standard Model
  effective field theory via Hilbert series methods}},''
  \href{http://dx.doi.org/10.1007/JHEP02(2016)081}{JHEP {\normalfont \bfseries
  02} (2016)  081}, \href{http://arxiv.org/abs/1510.00372}{{\normalfont
  \ttfamily arXiv:1510.00372}}.

\bibitem{henning20172}
B.~Henning, X.~Lu, T.~Melia, and H.~Murayama, ``{\em {2, 84, 30, 993, 560,
  15456, 11962, 261485, ...: Higher dimension operators in the SM EFT}},''
  \href{http://dx.doi.org/10.1007/JHEP08(2017)016}{JHEP {\normalfont \bfseries
  08} (2017)  016}, \href{http://arxiv.org/abs/1512.03433}{{\normalfont
  \ttfamily arXiv:1512.03433}}. [Erratum: JHEP 09, 019 (2019)].

\bibitem{Liao:2016qyd}
Y.~Liao and X.-D. Ma, ``{\em {Operators up to Dimension Seven in Standard Model
  Effective Field Theory Extended with Sterile Neutrinos}},''
  \href{http://dx.doi.org/10.1103/PhysRevD.96.015012}{Phys. Rev. D {\normalfont
  \bfseries 96} (2017) no.~1, 015012},
  \href{http://arxiv.org/abs/1612.04527}{{\normalfont \ttfamily
  arXiv:1612.04527}}.

\bibitem{Merlo:2016prs}
L.~Merlo, S.~Saa, and M.~Sacrist\'an-Barbero, ``{\em {Baryon Non-Invariant
  Couplings in Higgs Effective Field Theory}},''
  \href{http://dx.doi.org/10.1140/epjc/s10052-017-4753-5}{Eur. Phys. J. C
  {\normalfont \bfseries 77} (2017) no.~3, 185},
  \href{http://arxiv.org/abs/1612.04832}{{\normalfont \ttfamily
  arXiv:1612.04832}}.

\bibitem{Kobach:2017xkw}
A.~Kobach and S.~Pal, ``{\em {Hilbert Series and Operator Basis for NRQED and
  NRQCD/HQET}},'' \href{http://dx.doi.org/10.1016/j.physletb.2017.06.026}{Phys.
  Lett. B {\normalfont \bfseries 772} (2017)  225--231},
  \href{http://arxiv.org/abs/1704.00008}{{\normalfont \ttfamily
  arXiv:1704.00008}}.

\bibitem{henning2017operator}
B.~Henning, X.~Lu, T.~Melia, and H.~Murayama, ``{\em {Operator bases,
  $S$-matrices, and their partition functions}},''
  \href{http://dx.doi.org/10.1007/JHEP10(2017)199}{JHEP {\normalfont \bfseries
  10} (2017)  199}, \href{http://arxiv.org/abs/1706.08520}{{\normalfont
  \ttfamily arXiv:1706.08520}}.

\bibitem{Anisha:2019nzx}
Anisha, S.~Das~Bakshi, J.~Chakrabortty, and S.~Prakash, ``{\em {Hilbert Series
  and Plethystics: Paving the path towards 2HDM- and MLRSM-EFT}},''
  \href{http://dx.doi.org/10.1007/JHEP09(2019)035}{JHEP {\normalfont \bfseries
  09} (2019)  035}, \href{http://arxiv.org/abs/1905.11047}{{\normalfont
  \ttfamily arXiv:1905.11047}}.

\bibitem{marinissen202083106786}
C.~B. Marinissen, R.~Rahn, and W.~J. Waalewijn, ``{\em {..., 83106786,
  114382724, 1509048322, 2343463290, 27410087742, ... efficient Hilbert series
  for effective theories}},''
  \href{http://dx.doi.org/10.1016/j.physletb.2020.135632}{Phys. Lett. B
  {\normalfont \bfseries 808} (2020)  135632},
  \href{http://arxiv.org/abs/2004.09521}{{\normalfont \ttfamily
  arXiv:2004.09521}}.

\bibitem{graf20212}
L.~Graf, B.~Henning, X.~Lu, T.~Melia, and H.~Murayama, ``{\em {2, 12, 117,
  1959, 45171, 1170086, \textellipsis{}: a Hilbert series for the QCD chiral
  Lagrangian}},'' \href{http://dx.doi.org/10.1007/JHEP01(2021)142}{JHEP
  {\normalfont \bfseries 01} (2021)  142},
  \href{http://arxiv.org/abs/2009.01239}{{\normalfont \ttfamily
  arXiv:2009.01239}}.

\bibitem{Yu:2021cco}
B.~Yu and S.~Zhou, ``{\em {Hilbert series for leptonic flavor invariants in the
  minimal seesaw model}},''
  \href{http://dx.doi.org/10.1007/JHEP10(2021)017}{JHEP {\normalfont \bfseries
  10} (2021)  017}, \href{http://arxiv.org/abs/2107.11928}{{\normalfont
  \ttfamily arXiv:2107.11928}}.

\bibitem{Yu:2022nxj}
B.~Yu and S.~Zhou, ``{\em {Spelling out leptonic CP violation in the language
  of invariant theory}},''
  \href{http://dx.doi.org/10.1103/PhysRevD.106.L051701}{Phys. Rev. D
  {\normalfont \bfseries 106} (2022) no.~5, L051701},
  \href{http://arxiv.org/abs/2203.00574}{{\normalfont \ttfamily
  arXiv:2203.00574}}.

\bibitem{Yu:2022ttm}
B.~Yu and S.~Zhou, ``{\em {CP violation and flavor invariants in the seesaw
  effective field theory}},''
  \href{http://dx.doi.org/10.1007/JHEP08(2022)017}{JHEP {\normalfont \bfseries
  08} (2022)  017}, \href{http://arxiv.org/abs/2203.10121}{{\normalfont
  \ttfamily arXiv:2203.10121}}.

\bibitem{Graf:2022rco}
L.~Gr\'af, B.~Henning, X.~Lu, T.~Melia, and H.~Murayama, ``{\em {Hilbert
  Series, the Higgs Mechanism, and HEFT}},''
  \href{http://arxiv.org/abs/2211.06275}{{\normalfont \ttfamily
  arXiv:2211.06275}}.

\bibitem{alonso2014renormalization}
R.~Alonso, E.~E. Jenkins, A.~V. Manohar, and M.~Trott, ``{\em {Renormalization
  Group Evolution of the Standard Model Dimension Six Operators III: Gauge
  Coupling Dependence and Phenomenology}},''
  \href{http://dx.doi.org/10.1007/JHEP04(2014)159}{JHEP {\normalfont \bfseries
  04} (2014)  159}, \href{http://arxiv.org/abs/1312.2014}{{\normalfont
  \ttfamily arXiv:1312.2014}}.

\bibitem{remmen2019consistency}
G.~N. Remmen and N.~L. Rodd, ``{\em {Consistency of the Standard Model
  Effective Field Theory}},''
  \href{http://dx.doi.org/10.1007/JHEP12(2019)032}{JHEP {\normalfont \bfseries
  12} (2019)  032}, \href{http://arxiv.org/abs/1908.09845}{{\normalfont
  \ttfamily arXiv:1908.09845}}.

\bibitem{bonnefoy2021beyond}
Q.~Bonnefoy, E.~Gendy, C.~Grojean, and J.~T. Ruderman, ``{\em {Beyond Jarlskog:
  699 invariants for CP violation in SMEFT}},''
  \href{http://dx.doi.org/10.1007/JHEP08(2022)032}{JHEP {\normalfont \bfseries
  08} (2022)  032}, \href{http://arxiv.org/abs/2112.03889}{{\normalfont
  \ttfamily arXiv:2112.03889}}.

\bibitem{arias2019discrete}
G.~Arias-Tamargo, A.~Bourget, A.~Pini, and D.~Rodr\'\i{}guez-G\'omez, ``{\em
  {Discrete gauge theories of charge conjugation}},''
  \href{http://dx.doi.org/10.1016/j.nuclphysb.2019.114721}{Nucl. Phys. B
  {\normalfont \bfseries 946} (2019)  114721},
  \href{http://arxiv.org/abs/1903.06662}{{\normalfont \ttfamily
  arXiv:1903.06662}}.

\bibitem{henning2022outer}
B.~Henning, X.~Lu, T.~Melia, and H.~Murayama, ``{\em {Outer automorphism
  anomalies}},'' \href{http://dx.doi.org/10.1007/JHEP02(2022)094}{JHEP
  {\normalfont \bfseries 02} (2022)  094},
  \href{http://arxiv.org/abs/2111.04728}{{\normalfont \ttfamily
  arXiv:2111.04728}}.

\bibitem{weinberg_1995}
S.~Weinberg, {\em QUANTUM FIELDS AND ANTIPARTICLES}, vol.~1,
  \href{http://dx.doi.org/10.1017/CBO9781139644167.007}{p.~191–258}.
\newblock Cambridge University Press, 1995.

\bibitem{grimus1997automorphisms}
W.~Grimus and M.~N. Rebelo, ``{\em {Automorphisms in gauge theories and the
  definition of CP and P}},''
  \href{http://dx.doi.org/10.1016/S0370-1573(96)00030-0}{Phys. Rept.
  {\normalfont \bfseries 281} (1997)  239--308},
  \href{http://arxiv.org/abs/hep-ph/9506272}{{\normalfont \ttfamily
  arXiv:hep-ph/9506272}}.

\bibitem{georgi1999lie}
H.~Georgi, {\em Lie algebras in particle physics: from isospin to unified
  theories}.
\newblock Taylor \& Francis, second~ed., 2000.

\bibitem{Bischer:2022rvf}
I.~Bischer, C.~D\"oring, and A.~Trautner, ``{\em {Telling compositeness at a
  distance with outer automorphisms and CP}},''
  \href{http://arxiv.org/abs/2212.07439}{{\normalfont \ttfamily
  arXiv:2212.07439}}.

\bibitem{Sun:2022aag}
H.~Sun, Y.-N. Wang, and J.-H. Yu, ``{\em {Hilbert Series and Operator Counting
  on the Higgs Effective Field Theory}},''
  \href{http://arxiv.org/abs/2211.11598}{{\normalfont \ttfamily
  arXiv:2211.11598}}.

\end{thebibliography}\endgroup

\end{document}